\documentclass[twocolumn]{jpsj2}
\title{Effect of Structural Parameters on Superconductivity in  Fluorine-Free LnFeAsO$_{1-y}$ (Ln=La,Nd)}

\author{Chul-Ho LEE, Akira IYO, Hiroshi EISAKI, Hijiri KITO, Maria Teresa FERNANDEZ-DIAZ$^1$, Toshimitsu ITO, Kunihiro KIHOU, Hirofumi MATSUHATA, Markus BRADEN$^2$, and Kazuyoshi YAMADA$^3$}

\inst{National Institute of Advanced Industrial Science and Technology, Tsukuba, Ibaraki 305-8568, Japan\\
$^{1}$Institut Laue-Langevin, BP 156, F-38042 Grenoble Cedex 9, France\\
$^{2}$$I\hspace{-.1em}I$. Physikalisches Institut, Universit\"at zu K\"oln, Z\"ulpicher Str. 77, D-50937 K\"oln, Germany\\
$^{3}$Institute for Materials Research, Tohoku University, Sendai 980-8577, Japan}

\abst{The crystal structure of LnFeAsO$_{1-y}$ (Ln = La, Nd) has
been studied by the powder neutron diffraction technique. The
superconducting phase diagram of NdFeAsO$_{1-y}$ is established
as a function of oxygen content which is determined by
Rietveld refinement. The small As-Fe bond length suggests that As
and Fe atoms are connected covalently.  FeAs$_4$-tetrahedrons
transform toward a regular shape with increasing oxygen
deficiency.  Superconducting transition temperatures seem to
attain maximum values for regular FeAs$_4$-tetrahedrons.}

\kword{oxypnictides, superconductivity, powder neutron diffraction, crystal structure analysis}

\begin{document}
\maketitle

The recent discovery of superconductivity in LaFeAsO$_{1-x}$F$_x$
with a transition temperature of $T_c$ = 26 K has triggered an
intense search for related new superconductors
\cite{Kamihara2008}. Immediately, it was found that $T_c$
increases up to 55 K by replacing La with Sm atoms
\cite{Ren-F-2008}. This is the highest $T_c$ besides that in
high-$T_c$ cuprates rendering oxypnictides a promising new
class of superconductors.

Because the $T_c$ of the oxypnictide superconductors is very high, it may
be difficult to explain the formation of Cooper pairs with the
conventional BCS theory. Replacing As by the lighter P results
in a strong suppression of the $T_c$ in LaFePO$_{1-x}$F$_x$
\cite{Kamihara2006}, whose origin also needs to be explained. To
elucidate the superconducting mechanism, information about the
crystal structure of these materials is very important.

The parent compounds of LnFeAsO (Ln = lanthanide) exhibit a
tetragonal structure with the space group P4/nmm at room
temperature\cite{Quebe-struc2000} (Fig. 1). Characteristically,
LnO and FeAs layers are stacked alternately.  Fe atoms are in a
four-fold coordination forming a FeAs$_4$-tetrahedron. Upon
doping, LnO layers provide carriers to FeAs layers where
superconductivity is expected to be induced. The function of the layers
appears clearly separated similarly to high-$T_c$
cuprates.

Very recently, it has been reported that the fluorine-free samples
LnFeAsO$_{1-y}$ show superconductivity with a maximum $T_c$ of 55 K,
as well \cite{Kito2008, Ren-O-2008}. Based on the nominal
composition, superconductivity appears to be induced in a wide
range of oxygen deficiencies,  0.3 $\leq$ y $\leq$ 0.8. Since the
oxygen deficiency is huge, the crystal structure could be
significantly modified compared with that of the parent LnFeAsO
compounds. Furthermore, one may doubt that the real oxygen
deficiency is as large as the nominal values. Neutron diffraction
measurements on the LnFeAsO$_{1-y}$ samples are required to
confirm the crystal structure as well as the oxygen content. In
this study, we report the relationship between superconductivity
and crystal structure in LnFeAsO$_{1-y}$ compounds.

\begin{figure}[tb]
\includegraphics[width=\columnwidth]{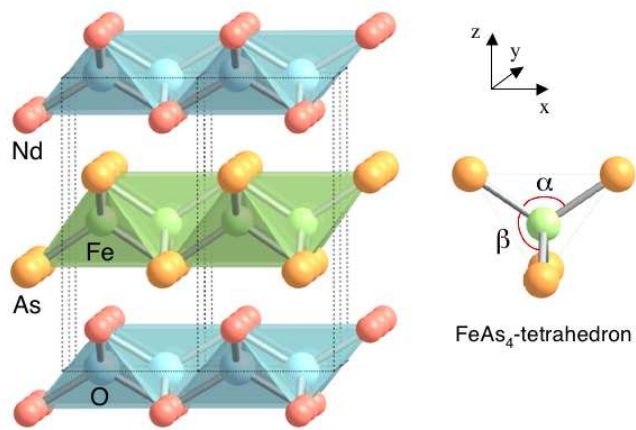}
\caption{\label{fig:crystal-structure} Crystal structure of
NdFeAsO.  Dashed lines show unit cells.  NdO and FeAs layers are
stacked alternately. FeAs$_4$ clusters in the FeAs layers form a
tetrahedral lattice.  The definitions of the two As-Fe-As bond angles
$\alpha$ and $\beta$ are illustrated on the right side with an FeAs$_4$-tetrahedron.}
\end{figure}

Polycrystalline samples of  LnFeAsO$_{1-y}$ (Ln=La,Nd) were synthesized
at high pressure and high temperature using a cubic-anvil
high-pressure apparatus.  Details of synthesis method are given
in ref. 5. The nominal oxygen deficiencies at the start of the synthesis was y =
0.15, 0.20, 0.40, and 0.30 in the 
NdFeAsO$_{1-y}$ samples labeled 1, 2, 3, and
4. For a LaFeAsO$_{1-y}$ sample, the nominal content is y
= 0.40 (sample 5). The oxygen composition of all the samples was
determined by the present Rietveld analysis, and we found that it
largely shifts towards higher oxidation (see Table 1). The amount
of samples used was about 0.5 g for each composition.

The $T_c$ of the LnFeAsO$_{1-y}$ (Ln=La,Nd) samples was measured
using a SQUID magnetometer under a magnetic field of 5 Oe after
zero field cooling from sufficiently above $T_c$ (Fig. 2). The
oxygen deficiency y indicated in Fig. 2 corresponds to that 
obtained from the Rietveld analysis of the neutron diffraction
patterns. The $T_c$ values of the La-based, y = 0.12, and of the
Nd-based, y = 0.08, 0.14, and 0.17, samples are 28, 35,
44, and 51 K, respectively.  On the other hand, the Nd-based
sample with y = 0.05 sample is not  superconducting.

\begin{figure}[tb]
\includegraphics[width=\columnwidth]{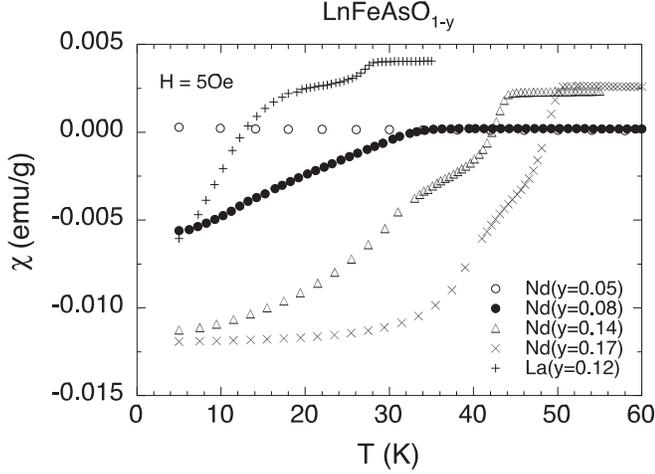}
\caption{\label{fig:SQUID} Shielding signals of LnFeAsO$_{1-y}$
measured under a magnetic field of $H$ = 5 Oe.  y was determined by Rietveld refinement.
The higher values in the normal states of LaFeAsO$_{1-y}$ (y = 0.12) and NdFeAsO$_{1-y}$ (y = 0.14, 0.17) could be due to slight impurities of iron compounds.}
\end{figure}

Neutron scattering measurement was carried out using the
high-resolution powder diffractometer D2B of the Institut
Laue-Langevin in Grenoble, France.  Incident neutron
wavelength was fixed at 1.594 \AA\ using a Ge monochromator.
Diffraction patterns were collected in the 2$\theta$ range of
9$^{\circ}$-160$^{\circ}$ at a constant step of 0.05$^{\circ}$
using a multidetector. Powder samples were placed in a
cylindrical vanadium can for measurements at room temperature.
These vanadium cans were mounted in a closed-cycle cryostat for
measurements at T = 10 K. The data were analyzed by the Rietveld
method using the Rietan program \cite{Izumi} for the refinement
of crystallographic structures.

\begin{figure}[tb]
\includegraphics[width=\columnwidth]{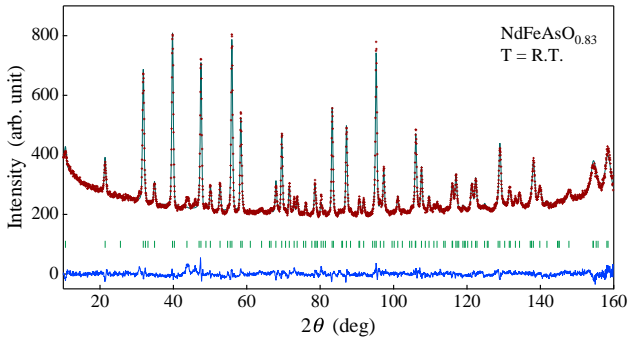}
\caption{\label{fig:Rietveld}  Typical observed (crosses) and calculated (solid lines) neutron powder diffraction patterns of NdFeAsO$_{1-y}$.  Vertical bars show the calculated positions of nuclear Bragg reflections.  The solid lines shown at the bottom of the figure indicate the differences between observations and calculations.}
\end{figure}
\begin{table}[structure-parameters]
\caption{Atomic parameters of LnFeAsO$_{1-y}$ (space group $P4/nmm$) determined by Rietveld refinements of neutron powder diffraction data.  $B$ is the isotropic atomic displacement parameter.}
\begin{center}
\begin{tabular}{c c l c c l l}\hline \hline
Atom    &   site &   occ. &   x &  y &\multicolumn{1}{c}{z}&\multicolumn{1}{c}{$B$ (\AA$^2$)} \\ \hline
\multicolumn{7}{l}{(a) sample 1 (non-super) T = R.T.}  \\
\multicolumn{7}{l}{$a$ = 3.96666(7)\AA,  $c$ = 8.5699(2)\AA, $R_{WP}$ = 2.98 \%}  \\
Nd&2c&1&1/4&1/4&0.1390(2)&0.46(4)\\
Fe&2b&1&3/4&1/4&0.5&0.57(4)\\
As&2c&1&1/4&1/4&0.6571(3)&0.50(5)\\
O&2a&0.95(1)&3/4&1/4&0&0.49(8)\\
\\
\multicolumn{7}{l}{(b) sample 2 ($T_c$ = 35 K) T = R.T.}  \\
\multicolumn{7}{l}{$a$ = 3.95940(6)\AA, $c$ = 8.5550(2)\AA, $R_{WP}$ = 2.66 \%}  \\
Nd&2c&1&1/4&1/4&0.1413(2)&0.34(3)\\
Fe&2b&1&3/4&1/4&0.5&0.42(3)\\
As&2c&1&1/4&1/4&0.6586(3)&0.53(4)\\
O&2a&0.920(9)&3/4&1/4&0&0.67(7)\\
\\
\multicolumn{7}{l}{(c) sample 3 ($T_c$ = 44 K) T = R.T.}  \\
\multicolumn{7}{l}{$a$ = 3.95365(7)\AA, $c$ = 8.5581(2)\AA, $R_{WP}$ = 3.25 \%}  \\
Nd&2c&1&1/4&1/4&0.1429(3)&0.23(4)\\
Fe&2b&1&3/4&1/4&0.5&0.37(4)\\
As&2c&1&1/4&1/4&0.6587(3)&0.39(5)\\
O&2a&0.86(1)&3/4&1/4&0&0.32(8)\\
\\
\multicolumn{7}{l}{(d) sample 4 ($T_c$ = 51 K) T = R.T.}  \\
\multicolumn{7}{l}{$a$ = 3.94755(7)\AA, $c$ = 8.5446(2)\AA, $R_{WP}$ = 2.84 \%}  \\
Nd&2c&1&1/4&1/4&0.1440(3)&0.42(4)\\
Fe&2b&1&3/4&1/4&0.5&0.40(3)\\
As&2c&1&1/4&1/4&0.6600(3)&0.41(5)\\
O&2a&0.83(1)&3/4&1/4&0&0.66(8)\\
\\
\multicolumn{7}{l}{(e) sample 4 ($T_c$ = 51 K) T = 10 K}  \\
\multicolumn{7}{l}{$a$ = 3.9423(1)\AA, $c$ = 8.5129(3)\AA, $R_{WP}$ = 4.07 \%}  \\
Nd&2c&1&1/4&1/4&0.1434(3)&0.46(6)\\
Fe&2b&1&3/4&1/4&0.5&0.49(5)\\
As&2c&1&1/4&1/4&0.6624(4)&0.36(6)\\
O&2a&0.83&3/4&1/4&0&0.48(9)\\
\\
\multicolumn{7}{l}{(f) sample 5 ($T_c$ = 28 K) T = R.T.}  \\
\multicolumn{7}{l}{$a$ = 4.02291(8)\AA, $c$ = 8.7121(2)\AA, $R_{WP}$ = 3.87 \%}  \\
La&2c&1&1/4&1/4&0.1453(3)&0.38(5)\\
Fe&2b&1&3/4&1/4&0.5&0.27(4)\\
As&2c&1&1/4&1/4&0.6527(4)&0.41(6)\\
O&2a&0.88(1)&3/4&1/4&0&0.54(9)\\
\hline \hline
\end{tabular}
\end{center}
\label{exp-list}
\end{table}
\begin{figure}[tb]
\includegraphics[width=\columnwidth]{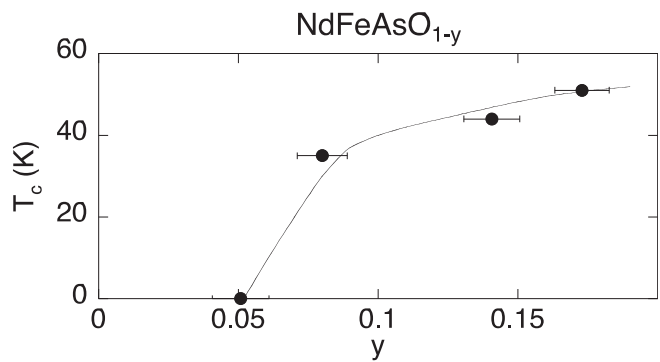}
\caption{\label{fig:Tc-dependence} $T_c$ vs oxygen deficiency y in NdFeAsO$_{1-y}$.}
\end{figure}
\begin{figure}[tb]
\begin{center}
\includegraphics[width=\columnwidth]{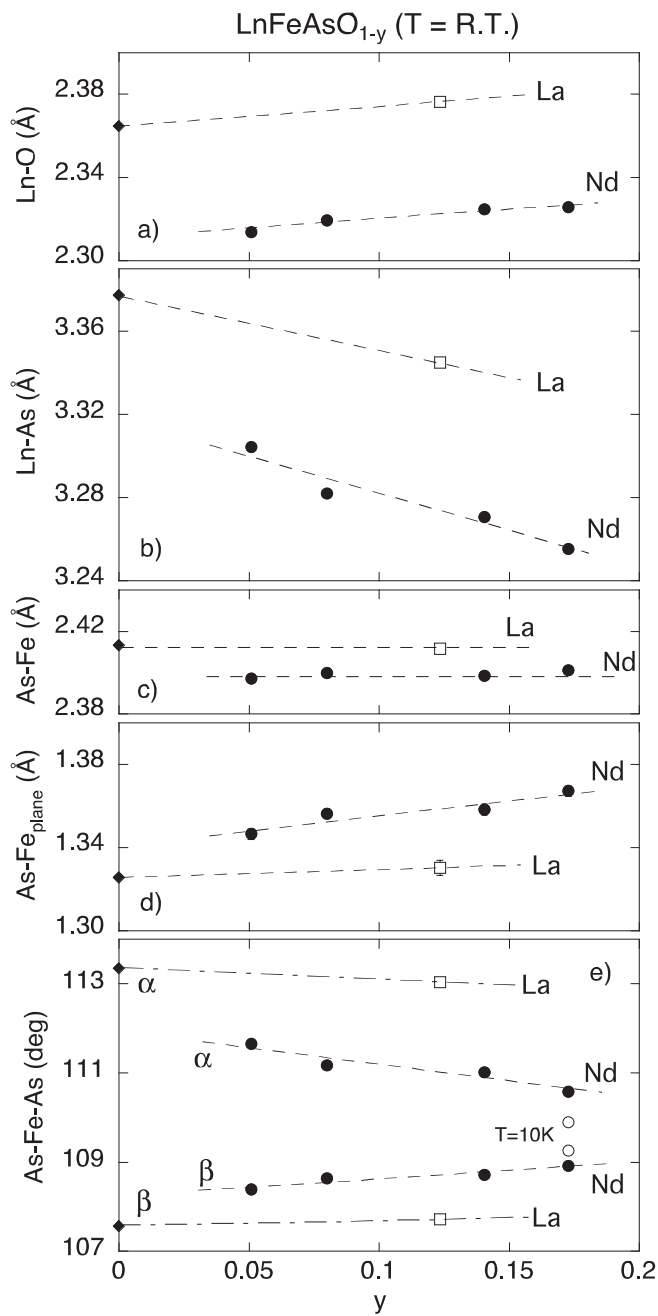}
\caption{\label{fig:parameters} Crystal structural parameters vs oxygen deficiency in LaFeAsO$_{1-y}$ at room temperature (open squares), NdFeAsO$_{1-y}$ at room temperature (closed circles), and NdFeAsO$_{1-y}$ at T = 10K (open circles).  The data of LaFeAsO (closed diamond) is cited from the literature \cite{Nomura-struc2008}.
The bond length of (a) Nd-O, (b) Nd-As, (c) As-Fe, (d) distance between the Fe plane and the As atom, and (e) As-Fe-As bond angles are depicted.
The definitions of the bond angles $\alpha$ and $\beta$ are illustrated in Fig. 1.}
\end{center}
\end{figure}
\begin{figure}[tb]
\includegraphics[width=\columnwidth]{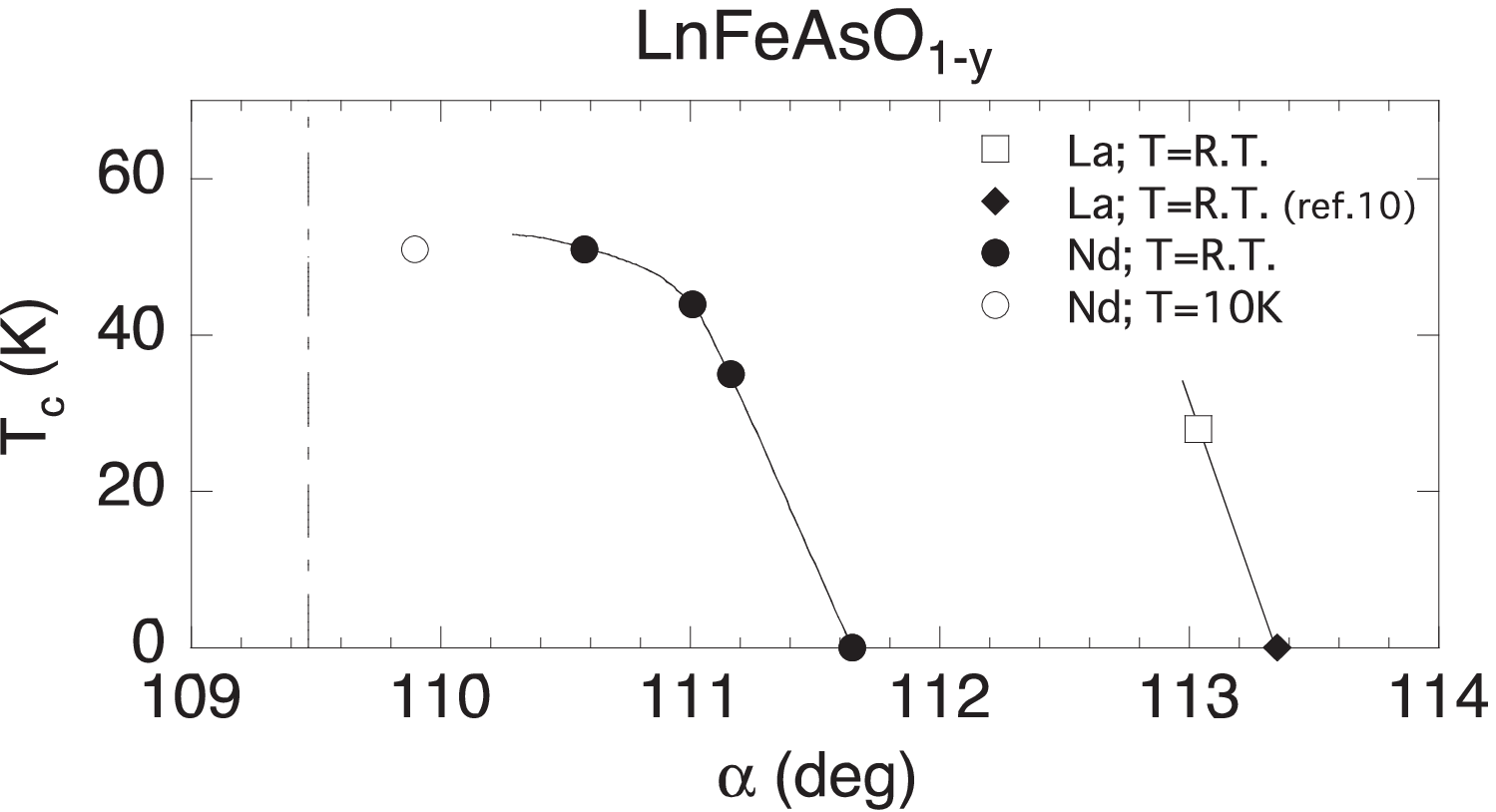}
\caption{\label{fig:Tc-tetrahedron} Relationship between $T_c$
and FeAs$_4$-tetrahedral distortion in LaFeAsO$_{1-y}$ at room
temperature (open square), NdFeAsO$_{1-y}$ at room temperature
(closed circles), and NdFeAsO$_{1-y}$ at T = 10K (open circle).
The value of LaFeAsO (closed diamond) was taken from ref. 10.  
The definition of As-Fe-As bond angle
$\alpha$ is illustrated in Fig. 1. The vertical dashed line
indicates the bond angle of a regular tetrahedron ($\alpha$ =
109.47$^{\circ}$).}
\end{figure}

Figure 3 shows a typical diffraction pattern of
NdFeAsO$_{0.83}$.  Small amounts of the impurity phases, LaAs,
NdAs, FeAs, La$_2$O$_3$, and Nd$_2$O$_3$ exist in the series of
measurements with a volume ratio of about 3\% at most. The space group
P4/nmm was assumed for all calculations; it has been confirmed recently
by the electron diffraction study on superconducting
NdFeAsO$_{1-y}$ \cite{Matsuhata2008}. Since the oxygen content of 
the synthesized samples can deviate from that of the starting
materials, the occupation of the oxygen site was varied in the
refinements. The obtained $R$ factors were in the range of 2.66 \%
$\leq$ $R_{WP}$ $\leq$ 4.07 \% (Table 1). The $R$ factor for
NdFeAsO$_{0.83}$ (sample 4) at T = 10 K is slightly larger owing to
some contamination in the diffraction pattern through the
cryostat. It turned out that the actual oxygen contents of the
samples are larger than the nominal (intended) values for all the
samples used in the current experiments. This is due to the
oxidation of the starting rare-earth elements.  

The superconducting phase diagram of NdFeAsO$_{1-y}$ versus the
oxygen deficiency y determined by Rietveld analysis is shown
in Fig. 4. The superconductivity appears above y = 0.05 and
attains a maximum $T_c$ value for NdFeAsO$_{1-y}$ compounds
around y = 0.17. The phase diagram is similar to that of the
recently reported fluorine-doped NdFeAsO$_{1-x}$F$_x$
\cite{Chen-F} which suggests an equivalent carrier doping level
when the y and x values are the same, although the amount of
induced carriers should be two times larger for oxygen deficiency.

The selected crystal structural parameters obtained from the Rietveld
analysis are shown in Fig. 5. For the non-doped LaFeAsO, the
parameters were extracted from the literature
\cite{Nomura-struc2008}, which is consistent with ref. 11.  Ln-O bond length increases with
increasing oxygen deficiency at the same rate in LaFeAsO$_{1-y}$
and NdFeAsO$_{1-y}$, possibly owing to the decrease in the number
of electrons in the O-planes.  The bond lengths may be
analyzed by the bond-valence-sum (BVS) formalism
\cite{Covalent2008,Shannon1976,brese}, in which each bond with
a distance $r$ contributes a valence $v=exp[{(d-r)\over0.37}]$ with
$d$ as an empirical parameter. In the non-superconducting
NdFeAsO$_{0.95}$, the four Nd-O bonds contribute 2.235 to the
Nd-BVS after correction for the incomplete oxygen occupation. In
contrast, the four Nd-As bonds, which are significantly longer,
contribute only 0.870 to the Nd-BVS which demonstrates the very
anisotropic bonding of the Nd coordination. The total Nd-BVS
is 3.104 in good agreement with the expected Nd valency.
The increase in Ln-O bond length with doping strongly reduces the
Nd-O contribution to the Nd-BVS to 1.888 for y=0.17. This
reduction is only partially compensated through the pronounced
shrinking of Nd-As distance with doping. For y=0.17, the
Nd-As bonds contribute 0.994 to the Nd-BVS resulting in a total
Nd-BVS value of 2.882, which is still close to the expected value. In
contrast, the BVS at the Fe site is doping-independent and always
far above the value expected for an ionic picture; Fe-BVS=3.522
for y=0.05 and 3.485 for y=0.17. The enhanced BVS
directly indicates that Fe-As bonds are too short compared with an
ionic picture and strongly suggests a covalent bonding similar to
the cuprates \cite{Covalent2008}.

The La and Nd series differ in doping dependence
of the thickness of the FeAs layer, which is determined by the
distance between the As sites and the plane through the Fe sites.
This As-Fe$_{plane}$ distance increases more rapidly in the
Nd-based compounds than in the La-based compounds. Related to
this, the As-Fe-As bond angles of the Nd-based compounds vary
more rapidly with increasing oxygen deficiency approaching a
regular FeAs$_4$ tetrahedron in which $\alpha$ and $\beta$ are 
109.47$^{\circ}$. Also upon cooling, the FeAs$_4$-tetrahedrons
become more regular.

\begin{figure}[tb]
\includegraphics[width=\columnwidth]{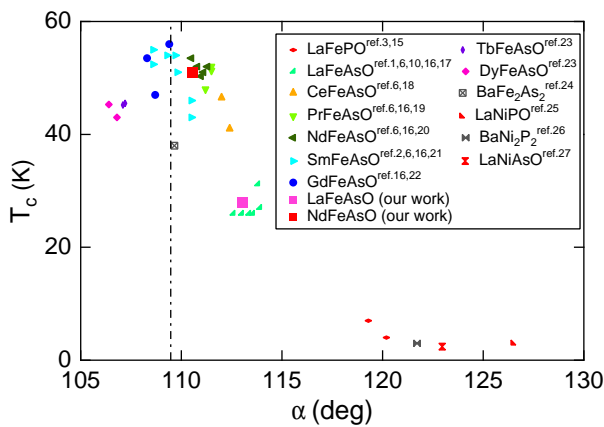}
\caption{\label{fig:bond-angle} $T_c$ vs As-Fe-As bond angle $\alpha$ for various pnictide superconductors.
Formulas of parent compositions of superconductors are depicted in the inset.
Crystal structure parameters of samples showing almost maximum $T_c$ in each system are selected.
The vertical dashed line indicates the bond angle of a regular tetrahedron ($\alpha$ = 109.47$^{\circ}$).}
\end{figure}

The relationship between $T_c$ and FeAs$_4$-tetrahedral
distortion is shown in Fig. 6. The tetrahedral distortion is
represented by the As-Fe-As bond angle. Clearly, $T_c$ increases
as the FeAs$_4$ coordination becomes a regular tetrahedron. This
suggests that the maximum $T_c$ values are attained when the
FeAs$_4$-lattices form a regular tetrahedron.  
This tendency could not change by lowering temperature, 
since they become closer to a regular tetrahedron at T = 10 K for the sample showing the highest $T_c$.  
To confirm this 
idea, we show in Fig. 7 As-Fe-As bond angle as a function of
$T_c$ in various pnictide superconductors
\cite{Kamihara2008,Ren-F-2008,Kamihara2006,Ren-O-2008,Nomura-struc2008,LaFePO,Iyo,LaFeAsO,CeFeAsO,PrFeAsO,NdFeAsO,SmFeAsO,GdFeAsO,Tb-DyFeAsO,BaFe2As2,LaNiPO,BaNi2P2,LaNiAsO}.  
The parameters of the samples showing almost maximum $T_c$ in each system are selected to eliminate the effect of carrier doping.  
The As-Fe-As bond angles in several LnFeAsO$_{1-y}$ compounds were
estimated from the lattice constants assuming a constant As-Fe
bond length of 2.40 \AA. This assumption is supported by the
present results which suggest that the rare-earth dependence of
As-Fe bond length is small.  This estimation can cause a bond-angle error of
about 1$^{\circ}$. Clearly, $T_c$ becomes maximum when
FeAs$_4$-lattices form a regular tetrahedron. This result
indicates a clear relationship between crystal structure and
superconductivity.

In summary, we have studied the crystal structure of
(La,Nd)FeAsO$_{1-y}$ by the neutron diffraction technique.  
Rietveld analysis revealed that the real oxygen content is largely
above the nominal composition. We present the superconducting
phase diagram of NdFeAsO$_{1-y}$ against the actual oxygen content.
FeAs$_4$-lattices were transformed toward a regular tetrahedron
accompanied by an increase in $T_c$ with increasing oxygen
deficiency y. It seems that $T_c$ becomes maximum when the
FeAs$_4$-lattices form a regular tetrahedron.

\begin{acknowledgments}
The authors would like to thank K. Kuroki, H. Aoki,  R. Arita, K.
Terakura, S. Ishibashi, I. Hase, N. Takeshita, K. Miyazawa, P. M.
Shirage, and R. Kumai for valuable discussions and the ILL for beam time allocation. This work was 
supported by Grants-in-Aid for Scientific Research on Priority
Areas and for Scientific Research B (No. 19340106) from MEXT,
Japan, and performed under the interuniversity cooperative
research program of the Institute for Materials Research, Tohoku
University. Work in Cologne was supported by the DFG through
SFB608.
\end{acknowledgments}

\end{document}